# Supercontinuum generation in the vacuum ultraviolet through dispersive-wave and soliton-plasma interaction in noble-gas-filled hollow-core photonic crystal fiber

A. Ermolov,* K. F. Mak, M. H. Frosz, J. C. Travers and P. St.J. Russell

Max Planck Institute for the Science of Light, Günther-Scharowsky-Str. 1, 91058 Erlangen, Germany

We report on the generation of a three-octave-wide supercontinuum extending from the vacuum ultraviolet (VUV) to the near-infrared, spanning at least 113 to 1000 nm (i.e., 11 to 1.2 eV), in He-filled hollow-core kagomé-style photonic crystal fiber. Numerical simulations confirm that the main mechanism is a novel and previously undiscovered interaction between dispersive-wave emission and plasma-induced blue-shifted soliton recompression around the fiber zero dispersion frequency. The VUV part of the supercontinuum, which modeling shows to be coherent and possess a simple phase structure, has sufficient bandwidth to support single-cycle pulses of 500 attosecond duration. We also demonstrate, in the same system, the generation of narrower-band VUV pulses, through dispersive-wave emission, tunable from 120 to 200 nm with efficiencies exceeding 1% and VUV pulse energies in excess of 50 nJ.

## I. INTRODUCTION

Vacuum-ultraviolet (VUV) light (100 to 200 nm) has a wide range of applications. Broadband coherent VUV sources could be used to produce sub-femtosecond pulses for precise—temporal, spectral and spatial—excitation of the electronic resonances of many atoms, molecules and solids, and form the basis for advanced time-resolved pump-probe spectroscopy [1-3] or the control of individual electronic wave-packets [4,5]. Techniques such as attosecond transient absorption spectroscopy [6], photoemission spectroscopy [7], and photoionization mass-spectroscopy [8], would directly benefit. Other applications include seeding free-electron lasers [9,10] and control of chemical reactions [11].

Existing spatially coherent sources in the VUV region include large-scale synchrotrons and free-electron lasers, or inefficient, and often elaborate, discrete frequency-conversion systems. Examples include second harmonic generation in exotic crystals, such as $SrB_4O_7$ [12], third harmonic generation [13] and four-wave mixing [8,14] in gases. These techniques usually produce VUV radiation with a narrow relative frequency bandwidth, and cannot be used to generate coherent sub-femtosecond pulses for the applications described above.

Here we make use of soliton dynamics in gas-filled kagomé-style photonic crystal fibers (kagomé-PCFs) [15-17], to efficiently generate tunable, coherent, ultrafast pulses from 120 nm to beyond 180 nm (11 to 7 eV). In addition, we show that in He-filled kagomé-PCF, this emission can evolve into a flat supercontinuum (SC) spanning from at least 113 to 1000 nm (11 to 1.2 eV)—over three octaves. This is the shortest wavelength VUV SC generated in any system. Although soliton dynamics in kagomé-PCF have been widely explored [16,17], showing, amongst many effects, a straightforward route to the generation of tunable deep-UV light [18,19], the mechanism demonstrated in this paper, based on a combination of resonant dispersive wave emission and soliton-plasma interactions, is completely novel.

## II. BACKGROUND

SC generation has been studied in glasses [20], gases [21] and liquids [22]. It is well established that in bulk materials the high-frequency SC edge scales with the frequencies of the ultraviolet electronic resonances of the medium, which are directly related to the ionization potential ($I_p$) [23]. This is primarily due to strong material dispersion [24]. Materials with the highest $I_p$, however, also have the lowest values of nonlinearity [25], thus in free space VUV SC generation requires very intense laser systems and filamentation, limiting the repetition rate and degrading the spatial coherence. With 33 GW few-cycle pump pulses in argon, the SC edge has reached ~210 nm at the 30 dB level [26]. Alternatively, the use of plasma defocusing of single-cycle pulses in Ne, yielded a UV SC extending from 150 to 300 nm [27]. Truly continuous extensions into the VUV have required TW pulses and spatially incoherent multi-filamentation; the shortest wavelength achieved in that regime is ~150 nm at the 40 dB level in Ar [28].

Solid-core PCFs [29] enhance the generation of broad and flat SC [30,31], by providing extended light-matter interaction lengths and anomalous (i.e. negative) group velocity dispersion (GVD) at common near-infrared pump wavelengths. Thus effects such as soliton-effect self-compression to few-cycle or even sub-cycle pulses [16,32,33], soliton fission [34–36] and resonant dispersive-wave emission at higher frequencies [18,19, 36-38] come into play. These processes have been harnessed to extend SC pumped by near-infrared laser pulses into the deep-UV. Glass-core optical fibers cannot, however, transmit in the VUV, the shortest wavelength SC edge achieved in solid-core PCF being

280 nm in silica-based fibers [39], and 200 nm in PCFs made from ZBLAN glass [40].

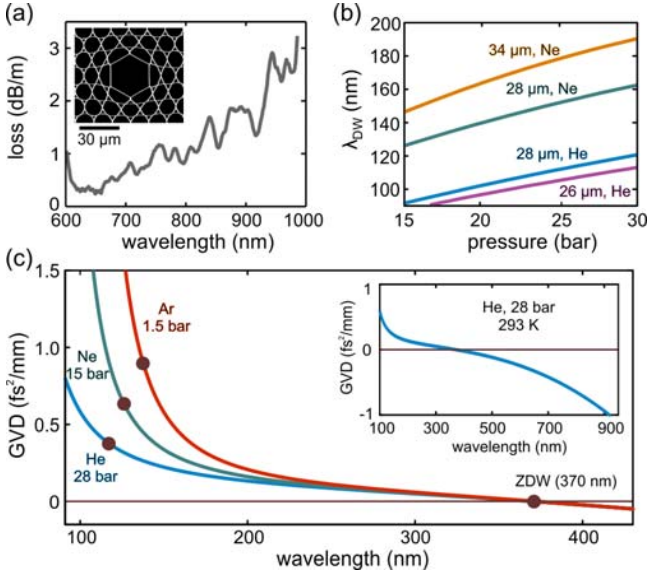

Fig. 1. (Color online) (a) Measured loss of kagomé-PCF with a core diameter of 28 μm. The inset shows a SEM of a typical kagomé-PCF. (b) Dependence of DW wavelength on pressure for various gases and fiber core diameters (calculated according to pulse duration − 35 fs, pulse energy − 4 μJ, 3.5 μJ, 5 μJ, 5 μJ for orange, green, blue and purple curves respectively). (c) Dispersion characteristics of the kagomé-PCF in (a), obtained for three different gases; the pressures were chosen to yield the same ZDW in each case. The brown dots in the vicinity of 150 nm denote the DW wavelength for each case. The inset shows the calculated dispersion characteristics from the VUV to the NIR when the 28 μm fiber is filled with 28 bar of He.

Gas-filled kagomé-PCF [15-17] (see scanning electron micrograph (SEM) in Fig. 1(a)) permits broadband low-loss transmission at core diameters of a few tens of μm. In such small cores the anomalous waveguide dispersion is large enough to balance the normal dispersion of the filling gas at pressures of a few bar. This makes it possible to operate with anomalous dispersion at pump wavelengths in the near-infrared (NIR), while benefiting from VUV transmission [41] and very high damage thresholds. Fig. 1(a) shows the transmission loss of a 28 μm kagomé-PCF from 600 to 1000 nm, and the inset of Fig 1(c) shows the dispersion from the VUV to near IR when it is filled with 28 bar of He. The zero dispersion wavelength (ZDW) is at 370 nm and the dispersion at 800 nm is –0.65 fs²/mm.

Three mechanisms for up-shifting the frequency of pump pulses at 800 nm in gas-filled kagomé-PCF have been previously reported: (i) efficient, coherent and tunable dispersive wave (DW) emission (also known as resonant or Cherenkov radiation in earlier publications) to the deep-UV and visible regions (180 to 550 nm) [18,19]; (b) the soliton self-frequency blue-shift, driven by photoionization [42–45]; and (c) impulsive Raman scattering [41]. The last of these made use of a hydrogen-filled kagomé-PCF, where impulsively excited Raman coherence resulted in strong spectral broadening into the VUV (124 nm at the 40 dB level) of DWs initially generated at ~180 nm. However, hydrogen has an $I_p$ similar to argon, so direct generation of light at wavelengths below 120 nm is unlikely, and in addition the VUV SC is not especially flat.

Here we report a new technique for generating a bright, flat VUV SC extending to below 113 nm, as well as extending high-energy wavelength-tunable VUV dispersive-wave emission down to 120 nm.

The transfer of radiation, from a self-compressed soliton pump pulse to a DW at a certain up-shifted frequency, is phase-matched by higher order dispersion [46–48]. The process can be understood as cascaded four-wave mixing [49]. Neglecting the nonlinear phase shift and dispersion terms higher than third order, the DW frequency $\omega_{dw}$ is given by:

$$\omega_{dw} = \omega_{sol} + 3|\beta_2|/\beta_3 = 3\omega_{zd} - 2\omega_{sol} \qquad (1)$$

where $\omega_{zd}$ is the frequency of the ZDW and $\omega_{sol}$ is the central frequency of the pump soliton; $\beta_2$ is the GVD, and $\beta_3$ is the third order dispersion, both at the pump frequency. This indicates that VUV light (< 200 nm) can be generated from a NIR pump pulse at 800 nm if the ZDW is in the UV region (<400 nm). According to Eq. (1), kagomé-PCFs filled with different gases will emit DWs at the same UV frequency $\omega_{dw}$ if the pressure is adjusted so that they share the same ZDW (or equivalently the same value of $|\beta_2|/\beta_3$). In practice this is not precisely true: at frequencies close to the electronic transitions, the group velocity dispersion (GVD) varies rapidly (Fig. 1(c)), so that higher order dispersion (i.e. $\beta_4$ and higher) alters the phase-matching condition in Eq. (1). The predicted wavelengths of the DWs, taking account of higher order dispersion, are plotted as a function of pressure in Fig. 1(b) for two different gases and three different core diameters. In keeping with the general trend in SC generation [23], the shortest VUV wavelengths can be reached only with the lightest (i.e. with the highest $I_p$) gases (Ne and He), resulting in tunability from 100 to 200 nm.

### III. EXPERIMENT

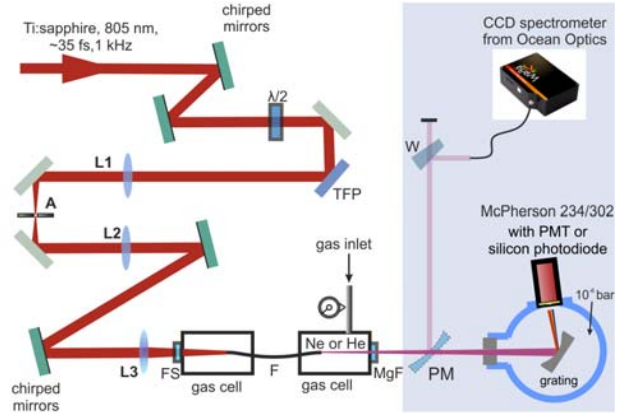

Fig. 2. (Color online) Schematic of experimental setup; the elements are described in the main text.

The experimental setup is shown in Fig. 2. Linearly polarized, 35 fs, pump pulses at 800 nm, and 1 kHz repetition rate, generated by a Ti:sapphire amplifier (Coherent, Legend Elite), were reflected by two pairs of chirped mirrors to compensate for the GVD induced by the subsequent spatial filter, achromatic focusing lens and input window (the second pair was placed right before the gas cell). A combination of thin-film polarizer (TFP) and half-wave plate (λ/2) allowed the input energy to be controlled from 100's nJ to 10's µJ. The spatial filter consisted of focusing (L1) and collimating (L2) lenses and a 100 µm aperture (A) between them. We mounted several kagomé-PCFs (F) with different core sizes between two gas-cells [16,17] and filled them with He and Ne gas at pressures of up to 30 bar. The gas cells were connected by plastic tubing, with the kagomé-PCF running through it. To deliver light in and out of the gas cells, fused silica (FS) and $MgF_2$ (MgF) windows respectively were sealed into the cell faces. The beam was focused onto the fiber tip by an achromatic lens (L3). The gas cells were evacuated and purged several times before being filled with the chosen gas at constant pressure. For VUV diagnostics the gas cell was connected to a McPherson vacuum spectrometer (234/302) via a bellows. The spectrometer was evacuated by a turbomolecular vacuum pump down to µbar pressure. The VUV detectors used were either a sodium salicylate scintillator and photo-multiplier tube (McPherson), or a silicon photodiode (Opto Diode Corporation), which had been absolutely calibrated between 90 and 220 nm using a synchrotron light source. To measure the entire spectrum including visible and near IR parts, the McPherson spectrometer was replaced by a parabolic mirror (PM) and fused silica wedge (W) to collimate and attenuate the beam coming out of the fiber and steer it to an Ocean Optics MayaPro CCD spectrometer. The measured spectrum was corrected for the spectrometer response and wavelength dependent diffraction from the output of the fiber.

The VUV spectra for Ne pressures of 26 and 21 bar are shown in Fig. 3a(i,ii), where a 34 µm diameter kagomé-PCF was used with 4 µJ pump energy. The results show that the DW frequency can be shifted into the VUV purely by pressure tuning (tuning out to 550 nm using Ne and heavier gases has been previously demonstrated [20]). Furthermore additional DWs at even shorter wavelength are also emitted into higher order modes [50], as for example seen in the red curve (marked – (i) in Fig. 3a), where the peak at 154 nm corresponds to a DW emitted into the $HE_{12}$ mode (confirmed by numerical modeling – black curve – (iv)). Reducing the core size to 28 µm and increasing the pressure to 21 bar shifts the ZDW further into the VUV, resulting in DW emission at 145 nm (Fig. 3a(iii)). We have performed the measurements of energy contained in these peaks. To do this the PMT of McPherson spectrometer was replaced by a silicon photodiode from Opto Diode Corporation (Fig.2). The input slit of the McPherson spectrometer was removed to allow the entire beam to enter (effectively utilizing the output of the fiber as the input slit) and the output slit was set up to have 1 nm resolution (verified using a mercury lamp). We corrected the measured spectrum using the known photodiode responsivity, grating efficiency and $MgF_2$ window transmission. To eliminate the effect of scattered light in the monochromator chamber the measurements were repeated with exactly the same parameters for each spectrum, but with the chamber flushed with air to absorb the VUV, or with a variety of glass and crystal filters. This signal was then subtracted from the original to avoid overestimation of the VUV power. The measured energy varied from 5 to 50 nJ in a range 110 – 200 nm depending on the fiber diameter, length and phase-matching conditions defined by pump energy, gas species and pressure. It corresponds to a conversion efficiency to the VUV of ~1%, significantly higher than obtained using the other approaches discussed in the introduction.

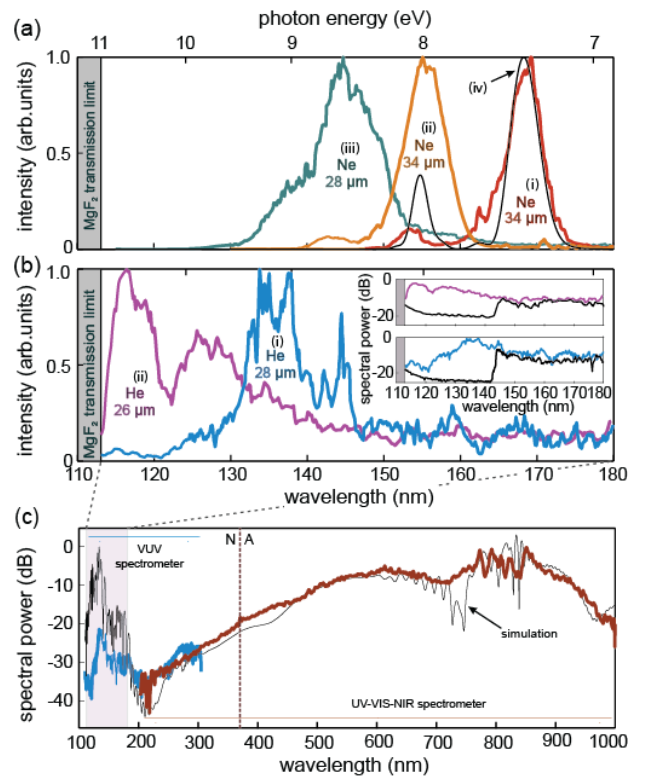

Fig. 3. (Color online) (a) Experimental spectra of coherent ultrashort pulses generated by DW emission in Ne-filled kagomé-PCF for parameters given in the text (each curve is individually normalized). The red and orange curves were taken at different gas pressures, all other parameters being fixed. The numerical simulation (solid black line marked (iv)) reproduces the experimental red curve and shows a small peak at 154 nm corresponding to DW emission into the $HE_{12}$ mode. (b) The VUV SC in He (linear scale) for core diameters of (i) 28 µm and (ii) 26 µm; (inset) the same data on a logarithmic scale, including the spectra recorded through a sapphire filter (black lines). (c) The full SC corresponding to (b)(i): the blue curve was measured using the VUV spectrometer and the brown curve using the UV-NIR spectrometer. The solid black line is the simulated spectrum. The dashed vertical line marks the ZDW (N = normal, A = anomalous GVD).

Switching to 28.2 bar He in the kagomé-PCF with core diameter 28 µm, and pumping with 5 µJ energy, the

DW peak can be shifted to 135 nm (Fig. 3b (i)). Using an even smaller core diameter (26 µm) with 28 bar He, the DW could be tuned down to 120 nm (Fig. 3b (ii)) for 3.5 µJ pump energy.

These He spectra are not discrete DW peaks, but in fact broad supercontinua extending from below 113 nm to beyond the range of the VUV spectrometer. Note that the spectra in Fig. 3b are displayed on a linear scale; the inset shows them on a logarithmic scale, emphasizing the extreme (<10 dB) flatness of these supercontinua across the VUV. The 113 nm short-wavelength cut-off is caused by the absorption edge of the $MgF_2$ window; the spectra probably extend to even shorter wavelengths. The measurement was repeated with a sapphire window placed in front of the spectrometer input slit, cutting the signal below 145 nm and confirming that it was real and not due to stray light (solid black lines in Fig 3b, inset). For the remainder of this article we consider the parameters of Fig. 3b(i). The full extent of the SC, which extends into the infrared, is plotted in Fig. 3c, showing that it is continuous from 113 nm to beyond 1000 nm with 35 dB flatness—a spectral range exceeding three octaves. This is the shortest wavelength edge of any continuously extending supercontinuum reported to date.

## IV. THEORY AND DISCUSSION

To understand the nonlinear dynamics, we numerically simulated the propagation of the pulse using a multimode, carrier-resolved, unidirectional propagation equation described in detail in [50]. We included the Kerr effect, self-focusing, ionization and plasma effects.

The nonlinear refractive index data was taken from [51], and the full modal dispersion were calculated using the Marcatili model for capillaries [52], which has previously been shown to be valid in the near-IR to deep-UV range [16]:

$$n_{nm}(\omega) = \sqrt{n_0^2(\omega) - \frac{u_{nm}^2}{k^2 a^2}} \qquad (2)$$

where $n_0$ is the refractive index of the gas, $u_{nm}$ is the $n$-th root of the $(m-1)$th order Bessel function of the first kind, $k = \omega/c$ and $a$ is the core radius (defined as the flat-to-flat distance for the hexagonal structure of the kagomé-PCF core). For the refractive index of neon, we used the Sellmeier equation given in [53]. To describe the propagation of light at very high frequencies in helium, we developed a modified Sellmeier equation from published refractive index data down to 90 nm [54]:

$$n_0^2(\lambda) - 1 = \rho \sum_{i=1}^{3} \frac{B_i \lambda^2}{\lambda^2 - C_i} \qquad (3)$$

where $\rho$ is the gas density relative to 1 bar at 273 K, $\lambda$ is the wavelength in micrometers, and the values of $B_i$ and $C_i$ are given in Table 1.

For the photoionization rate we used the PPT model [55], modified with the ADK coefficients [56].

Our model for plasma polarization closely follows that reported in [57].

Table 1. Coefficients of Eq. (3) for helium at 1 bar and 273 K

| | |
|---|---|
| $B_1 = 2.16463 \times 10^{-5}$ | $C_1 = -6.8077 \times 10^{-4}$ |
| $B_2 = 2.10561 \times 10^{-7}$ | $C_2 = 5.13251 \times 10^{-3}$ |
| $B_3 = 4.75093 \times 10^{-5}$ | $C_3 = 3.18621 \times 10^{-3}$ |

The initial conditions for the numerical simulations were obtained from FROG traces of the pump laser pulses. The simulations show excellent agreement with experiment. Two examples are provided by the full black curves in Fig. 3a and 3c, demonstrating agreement both for discrete DW peaks using Ne (Fig. 3a), and for the full VUV-NIR SC (Fig. 3c). In the second case, the discrepancies in the VUV—which we note are only in terms of spectral power, and not spectral features—are most likely caused by fiber loss acquired during in propagation subsequent to the VUV generation point, which is unknown in this spectral region.

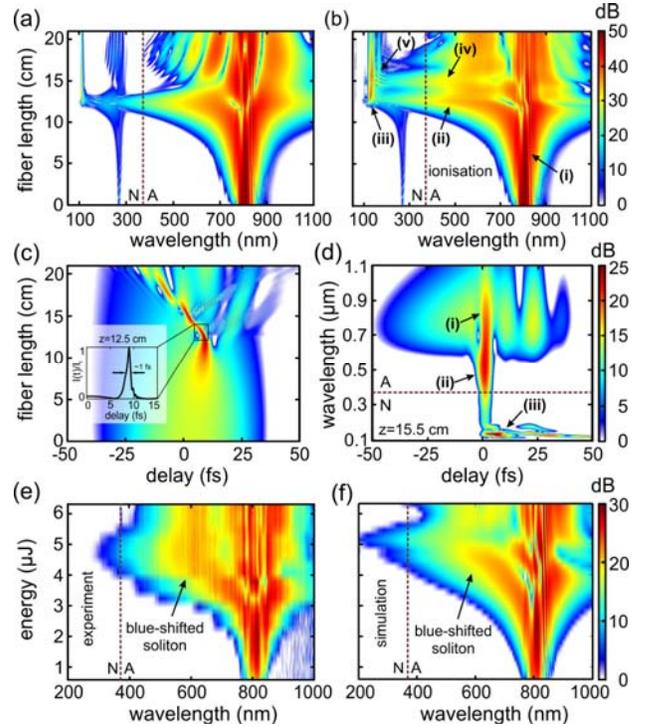

Fig. 4. (Color online) (a) Simulated spectral evolution for parameters given in the text, without ionization. (b) Same as (a) but with ionization. (c) Temporal evolution corresponding to (b); the inset shows the pulse profile at the point of maximum compression. (d) XFROG trace of the simulations in (b,c) at 15.5 cm. (e) Experimental and (f) numerical spectral evolution as a function of pump energy at the 16 cm point. The dashed lines in (a,b,d,e,f) mark the ZDW (N = normal, A = anomalous GVD).

Figure 4 shows the results of detailed numerical simulations for 28.2 bar He in a 28 µm diameter kagomé-PCF pumped with 5 µJ (the case in Figs. 3b (i) and 3c). Over the first 12.5 cm of propagation the pump pulse, corresponding to soliton order $N = 4.5$, undergoes self-compression [58]. In the spectral domain this initially results in approximately symmetric broadening due to self-phase modulation (marked (i) on Fig. 4b) that is continuously phase-compensated by anomalous dispersion. The resulting temporal self-compression is clearly evident in Fig. 4c, showing that the initial 35 fs pump pulse is compressed to a sub-cycle duration of less than 1 fs. This extreme compression is aided by self-steepening, leading to an optical shock [59]. The resulting blue-enhanced spectrum seeds the high-frequency DW emission, significantly increasing the conversion efficiency [18,60]. This is especially important for the frequency up-conversion into the VUV. While we do not directly measure the extremely short self-compressed pulse duration, we can confidently infer that it is produced within the fiber because it is essential for VUV DW generation, and is also supported by rigorous numerical modeling.

To emit a DW in the far-VUV (~120 nm), the core size and pressure were reduced so that the ZDW shifted to shorter wavelength. This causes the perturbation due to higher-order dispersion ($\beta_3/|\beta_2|$) at the pump wavelength (800 nm) to weaken. As a result, soliton-effect compression is less disturbed and shorter self-compressed pulse durations can be reached. Reducing the pressure also diminishes the nonlinearity so that more energy is required for pulse self-compression. For both of these reasons the self-compressed peak intensity increases such that it can be sufficient to partially ionize the gas [42,44]. In kagomé-PCF, due to the anomalous dispersion, this leads to a soliton self-frequency blue-shift [43–45] by over several hundred nanometers. This mechanism is crucial for enhancing the VUV emission. Fig. 4b shows the characteristic spectral signature of a blue-shifted soliton (marked (ii)), and the XFROG spectrogram in Fig. 4d confirms this (Fig. 4d – marked (ii)). This leads to a much stronger VUV signal (marked iii in Fig. 4b, 4d) compared to the case when ionization is switched off (compare Fig 4a to Fig. 4b), because the pump soliton has a stronger overlap with the DW frequency [61]. The emission is also at a longer wavelength (130 nm as opposed to 115 nm), in agreement with Eq. 1, because the pump soliton has shifted to a higher frequency.

Fig. 4e&f show the experimentally measured and numerically simulated evolution of the blue-shifting solitons with increasing pulse energy (1 to 6 μJ) for the parameters in Figs. 3b(i) and 3c. Experiment and simulation agree very well, the blue-shifting soliton extending down to 400 nm, where the shift is cancelled due to its encountering the ZDW. This cancellation, analogous to the cancellation of the Raman soliton self-frequency shift at long wavelengths in solid-core fibers [62], is caused by the emission of DWs in the normal dispersion region at higher frequencies. As the DW is emitted, the soliton recoils due to energy conservation—the recoil can be described as the Stokes side of a cascaded four-wave-mixing process [49]. From Fig. 4b we see that the soliton can subsequently reshape, blue-shift again (Fig. 4b-iv), and emit further DWs (Fig. 4b-v). This is confirmed by further detailed analysis of the XFROG traces. In this way the soliton 'bounces' several times off the zero dispersion point, each time emitting a DW in the vacuum ultraviolet. The effect of multiple DW emission is to significantly broaden the emitted spectral width. Since the role of plasma increases when phase-matching moves to shorter wavelength DWs, the broadening effect of the bouncing solitons explains the clear trend observable in Fig. 3a, b of increasing DW bandwidth as the DW peak is emitted at shorter wavelengths (Fig. 3a,b).

In the extreme case, the strong interaction between plasma-induced blue-shifted soliton recompression and "bouncing" off the ZDW and the emission of multiple DWs in the fundamental mode, in addition with emission into a few higher order modes [50], is a new mechanism capable of forming a continuous SC extending from below 113 nm to beyond 1000 nm, as shown in the experimental results of Fig. 3c. Note also that Fig. 3 and 4 represents the numerical data on a wavelength scale, for consistency with experimental measurements, however, the same propagation plot as in Fig 4b is shown in Fig. 5b, but on a frequency scale, which illuminates the VUV supercontinuum much more clearly.

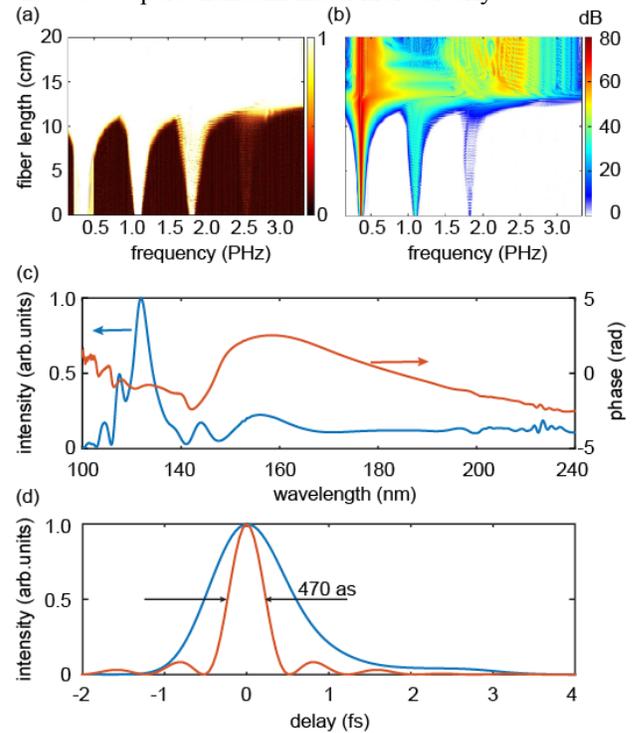

Fig. 5. (Color online) (a), (b) Simulated complex degree of first-order coherence and corresponding spectral propagation plot. The parameters are the same as in Fig 3b(i) and Fig 4b but the evolution is represented on a frequency scale. (c) The spectral intensity and phase of the VUV part of the supercontinuum from Fig 4b. (d) The time-domain intensity of the VUV part of the supercontinuum from Fig 4b. The blue curve is the uncompressed pulse, the orange curve is the pulse after phase-compensation.

Note that the SC emission remains temporally and spatially coherent, and is predominantly in the fundamental mode—although the first few $HE_{1n}$ modes also play a role, their spatial variation is weak and the spectrum is almost perfectly spatially homogeneous. A series of numerical simulations with different quantum and laser noise seeds, following [63], can be used to estimate the complex degree of first-order coherence, which can be simplified to

$$\left|g_{pq}^{(1)}(\omega)\right| = \left|\frac{\left\langle E_p^*(\omega)E_q(\omega)\right\rangle}{\left\langle \left|E_p(\omega)\right|^2\right\rangle}\right| \quad (4)$$

where the angle brackets denote an ensemble average over the independent simulations. If the electric fields had perfectly equal phase and intensity from shot to shot (i.e. are coherent), then the value of $|g_{pq}|$ would be 1, whereas a value of 0 indicates completely random phases between different laser shots. Fig. 5b and 5a shows the spectrum and the value $|g_{pq}|$ for propagation corresponding to Fig. 3b(i). The whole continuum is fully coherent. The VUV part, which has a simple phase structure, as shown in Fig. 5c, corresponds to a pulse of duration ~1 fs at the fiber output, even without any additional pulse compression, suggesting that this source could be used as a source of tunable few-femtosecond pulses in the VUV. If the phase is fully compensated, the duration reduced to 500 attoseconds (Fig. 5d).

## V. SUMMARY

We have demonstrated the generation of ultrashort VUV dispersive-waves in He and Ne-filled hollow-core kagomé-style photonic crystal fiber. The emission is tunable from 120 to 200 nm with efficiencies exceeding 1% and VUV pulse energies in excess of 50 nJ. We have also reported the generation of a three-octave-wide supercontinuum extending from the vacuum ultraviolet (VUV) to the near-infrared, spanning at least 113 to 1000 nm (i.e., 11 to 1.2 eV). Numerical simulations have demonstrated that the supercontinuum arises from an interaction between dispersive-wave emission, soliton recoil, recompression and plasma-induced soliton-blue-shift, causing the solitons to 'bounce' from the fiber zero dispersion frequency and emit multiple VUV dispersive-waves. Modeling also showed that the supercontinuum should be fully coherent. The VUV part possesses a simple phase structure having duration of ~500 attoseconds if the phase is fully compensated.